\documentstyle[11pt]{article}                   % scritto per latex
\hoffset=  - 1.2 true cm
\voffset=  - 1.4 true cm
\let\np\noindent
\long\def\comment#1{}

\def\csbul{{\tiny$\bullet$}\kern 0.2em}
\def\csd#1#2{\hbox{$#1$}\,\hbox{$\rm #2$}}
\def\cstimes{\cdot}
\def\cp{\ .}
\def\cv{\ ,\ } 

\makeatletter

\def\@cite#1{{\ [#1]}}

\makeatother

\def\au[#1]#2/{\bibitem{#1}{\sc #2},}       % autore
\def\isa[#1]{\bibitem{#1}}                  % item senza autore, cioe` senza
\def\asi #1/{{\sc{#1}},}          % autore senza item

\def\ti #1/{{\sl #1,}}          % titolo
\def\bt #1/{{\sl #1}}           % booktitle
\def\jo #1/{#1}                 % journal
\def\vo #1/{{{\bf #1},}}         % volume
\def\pu #1/{#1,}                 % publisher
\def\puend #1/{#1.}              % publisher at the end - con un punto
\def\olo #1{$#1$}                %
\def\yr #1/{{(\olo{#1})},}          % year
\def\yrend #1/{{(\olo{#1}).}}       % year at the end - con un punto
\def\pg #1/{{\olo{#1}}}        % pagina senza punto
\def\pgend#1/{{\olo{#1}}.}     % pagina alla fine - con punto
\def\pp #1-#2/{{\olo{#1}}\lower 0.2ex\hbox{--}{}{\olo{#2}}}
\def\ppend #1-#2/{{\olo{#1}}\lower 0.2ex\hbox{--}{}{\olo{#2}}.}
\def\and{{\it\&\/} }

\def\ii#1{{\it #1}\index{#1}}

\def\zz{@{\hspace{0em}}}  % per le tavole, per fare uno spazio zero
\def\zzs{@{\hspace{2pt}}}  % per le tavole, per fare uno spazio piccolo
\def\cstabhlineup{\rule{0mm}{3.0ex}}
\def\cstabhlinedown{\rule[-1.5ex]{0mm}{2ex}} 
\begin{document}

gr-qc/9610066

% Submitted to Physical Review D15

\vspace*{30mm}
\centerline{\Large\bf Does matter differ from vacuum?}

\bigskip
\bigskip
\bigskip

\centerline{Christoph Schiller$^{*}$} 
\medskip 
\centerline{Service de Chimie-Physique, CP 231} 
\centerline{Universit\'e Libre de Bruxelles} 
\centerline{Boulevard du Triomphe} 
\centerline{1050 Bruxelles, Belgium} 

\bigskip
\bigskip
\bigskip
\bigskip
\bigskip
\bigskip

%-----------------------------------------------------------------------------
\centerline{\bf Abstract}

\bigskip

\noindent 
A structured collection of thought provoking conclusions about space and time
is given. Using only the Compton  wavelength $\lambda= \hbar / m c$  and the
Schwarzschild radius $r_s=2 G m/ c^2$, it is argued that neither the
continuity of space-time nor the concepts of space-point, instant, or point
particle have experimental backing at high energies. It is then deduced that
Lorentz, gauge, and discrete symmetries are not precisely fulfilled in nature.
In the same way, using a new and simple {\it Gedankenexperiment}, it is
found that at Planck energies, vacuum is fundamentally indistinguishable from
radiation and from matter. Some consequences for supersymmetry, duality, and
unification are presented.

\bigskip
\bigskip
\bigskip

\centerline{PACS numbers: 03.65.Bz, 04.20.Cv}

%-----------------------------------------------------------------------------
%-----------------------------------------------------------------------------
%-----------------------------------------------------------------------------
\newpage \subsection{The limits of textbook physics}

Matter and space-time are the fundamental entities of our description of
nature.  Why do particles have the masses they have? The quantum theory of the
electromagnetic, weak and strong nuclear forces cannot answer the question. 
Why does space-time have three plus one dimensions?  The generally accepted
theory of the structure of space-time, general relativity, cannot explain this
simple observation (and in fact does not even allow to ask this question). One
concludes that the description of nature by these theories is still {\it
incomplete}.  

The search for a more complete, unified  description of nature is also driven
by another, even more compelling  motivation:  quantum mechanics and general
relativity {\it contradict\/} each other; they describe  the same phenomena in
incompatible ways. This well-known fact surfaces in many different
ways.\cite{technicalities}

There is a simple, but not well known way to state the origin of the
contradiction between general relativity and quantum mechanics. 
Both theories describe motion using objects, made of particles, and
space-time. Let us see how these two concepts are defined.

A particle --
and in general any object -- is defined as a conserved, localisable entity
which can move. (The etymology
of the term `object' is connected to this fact.) In other words, a particle 
is a small entity with conserved
mass, charge etc., which can vary its position with time.

At the same time, in every physics text\cite{timedef} time is
defined with the help of moving objects, usually called `clocks', or with the
help of moving particles, such as those emitted by light sources. Similarly,
also the length unit is defined with objects, be it a oldfashioned
ruler, or nowadays with help of the motion
of light, which is a collection of moving particles as well. 

The rest of 
physics has sharpened the definitions of particles and space-time. In quantum
mechanics one assumes space-time given (it is included as a symmetry of the
Hamiltonian), and one studies in all detail the properties and the motion of
particles from it, both for matter and radiation. In general relativity and
especially in cosmology, the opposite path is taken: one assumes that the
properties of matter and radiation are given, {\it e.g.}  via their equations
of state,
 and one describes in detail the space-time that follows from them, in
particular its curvature.  But one fact remains unchanged throughout all these
advances in standard textbook physics: {\it the two concepts of
particles and of space-time are defined with the help of each other.}  To avoid
the contradiction between quantum mechanics and general relativity and to try
to eliminate their joint incompleteness  requires  eliminating this 
circular definition. As argued in the following, this necessitates a radical 
change in our description of nature, and in particular about the continuity of
space-time.

For a long time, the contradictions in the two descriptions of nature were
avoided by keeping them separate: one often hears the statement that  quantum
mechanics is valid at small dimensions, and the other, general relativity, is
valid at large dimensions. But this artificial separation is not justified and
obviously prevents the solution of the problem. The situation resembles the
well-known drawing by M.C. Escher, where two hands, each holding a pencil, seem
 to draw each other. If one takes one hand as a symbol for space-time, the
other as a symbol for particles, and the act of drawing as  a symbol for the
act of defining, one has a description of standard, present day  physics. The
apparent contradiction is solved when one recognizes that both concepts (both
hands) result from a hidden third concept (a third hand) from which the other
two originate. In the picture, this third entity is the hand of the painter.

In the case of space-time and matter, the search for this underlying concept
is presently making renewed progress.\cite{ashtall,sstall,maggall,ishamfacie}
 The required conceptual changes are so dramatic that they should be of
interest to anybody who has an interest in physics. Some of the issues are
presented here. The most effective way to study these changes is to focus in
detail on that domain where the contradiction between the two standard
theories becomes most dramatic, and where both theories are  necessary at the
same time. That domain is given by the following well-known argument.

%-----------------------------------------------------------------------------
\subsection{Planck scales}

Both general relativity and quantum mechanics are successful theories for the
description of nature. Each of them provides a criterion to determine when
Galilean physics is not applicable any more. (In the following, we use the
terms  `vacuum' and `empty space-time' interchangeably.)

General relativity shows that  it is necessary to take into account the
curvature of space-time whenever one approaches an object of mass
$m$ to distances of the order of the Schwarzschild radius
$r_{\rm S}$, given by
\begin{equation}
  r_{\rm S} =  2 G m / c^2 \cp
\label{SCHW}
\end{equation}
 Approaching the Schwarzschild radius of an object, the difference between
general relativity and the classical $1/r^2$ description of gravity becomes
larger and larger. For example,  the barely measurable  gravitational 
deflection of light by the sun is due to an approach to $2.4
\times 10^{5}$ times the Schwarzschild radius of the sun.\cite{msw, wein}  In
general however, one is forced to stay away from objects an even larger
multiple of the Schwarzschild radius, except in the vicinity of neutron stars,
as shown in table~\ref{tsch}; for this reason, general relativity is not
necessary in everyday life. (An object smaller than its own Schwarzschild
radius is called a \ii{black hole}. Following general relativity, no signals
from the inside of the Schwarzschild radius can reach the outside world; hence
the name `black hole'.\cite{bhsearch})

Similarly, quantum mechanics shows that Galilean physics must be
abandoned and quantum effects must be taken into account whenever one
approaches an object at distances of the order of the  Compton wavelength
$\lambda_{\rm C}$, which is given by
\begin{equation}
    \lambda_{\rm C} =  { \hbar \over m c } \cp
\end{equation}
  Of course, this length only plays a role if the object itself is smaller
than its own Compton wavelength. At these dimensions one observes relativistic
quantum effects, such as particle-antiparticle creation and annihilation. 
Table~\ref{tsch} shows that the ratio $d/\lambda_{\rm C}$ is near or  smaller
than 1 only in the microscopic world, so that such  effects are not observed
in everyday life; therefore 
 one does not need quantum field theory to describe common observations.

Taking these two results together, the situations which   require the combined
concepts of quantum field theory and of general relativity are  those in which
both conditions are satisfied simultaneously.  The necessary approach distance
is calculated by setting $r_{\rm S} =  2\lambda_{\rm C}$ (the factor 2 is
introduced for simplicity). One finds that this is the case when lengths or
times are of the order of
\begin{equation}
\begin{array}{lcl\zz l}
 l_{\rm Pl}=  \sqrt{\hbar G/c^3}
    &= & \csd{1.6 \cdot 10^{-35}}{m} & \hbox{, the \ii{Planck length,} }\\
 t_{\rm Pl}=  \sqrt{\hbar G/c^5}
    &= & \csd{5.4 \cdot 10^{-44}}{s} & \hbox{, the \ii{Planck time.} }  \\
\end{array}
\end{equation}  
 Whenever one approaches objects at these dimensions, general relativity and
quantum mechanics both play a role; at these scales effects of  {\it quantum
gravity} appear. The values of the Planck dimensions being extremely small,
this level of sophistication is not necessary in everyday life, neither in
astronomy nor in present day microphysics. 

However, this sophistication is necessary to understand why the universe is the
way it is. The questions mentioned at the beginning -- why do we live in three
dimensions, why is the proton 1834 times heavier than the electron -- need for
their answer a precise and complete description of nature. The contradictions
between quantum mechanics and general relativity make the search for these
answers impossible. On the other hand, the unified theory, describing
quantum gravity, is not yet finished; but a few glimpses on its implications
can already taken at the present stage.

Note that the Planck scales are also the {\it only} domain of nature where
quantum mechanics and general relativity come together; therefore they 
provide the only possible starting point for the following discussion.
Planck\cite{planck} was interested in the Planck units mainly as natural
units of measurement, and that is the way he called them.  However, their
importance in nature is much more pervasive, as will be seen now.

%-----------------------------------------------------------------------------
\subsection{Farewell to instants of time}

\np The\label{timeaway} difficulties arising at Planck dimensions appear when 
one investigates the properties of clocks and meter bars. Is it possible to
construct a clock\index{clocks and Planck time} which is able to measure time
intervals shorter than the Planck time? Surprisingly, the answer is
no\cite{notea,targ}, even though in the relation $\Delta E \Delta t
\geq \hbar$ it seems that by making $\Delta E$ arbitrary large, one can make
$\Delta t$ arbitrary small.

Any clock is a device with some moving parts; such parts can be mechanical
wheels,  matter particles in motion, changing electrodynamic fields -- {\it
e.g.} flying photons \hbox{--,} decaying radioactive particles, etc.  For each
moving component of a clock, {\it e.g.} the two hands of the dial, the
uncertainty principle applies. As  has been discussed in many
occasions\cite{heis, kennard} and most clearly by Raymer\cite{uncert}, the
uncertainty relation for two non-commuting variables describes two different,
but related situations: it makes a statement  about standard {\it deviations}
of {\it separate} measurements on {\it many} identical systems, and  it
describes the measurement {\it  precision} for a {\it joint} measurement on  a
{\it single} system. Throughout this article, only the second viewpoint is
used. 

Now, in any clock, one needs to know both the time and the energy of each
hand,  since otherwise it would not be a classical system, {\it i.e.}  it
would not be a recording device. One therefore needs the joint knowledge of
non-commuting variables for each moving component of the clock; we are
interested in  the component with  the largest time uncertainty, {\it i.e.} 
imprecision of  time measurement
$\Delta t$.  It is evident that the smallest time interval
$\delta t$ which can be measured by a clock is always larger than the time
imprecision
$\Delta t$ due to the uncertainty relation for its moving components. Thus one
has
\begin{equation}
 \delta t \ge \Delta t \ge {\hbar \over \Delta E}
\end{equation} 
 where $\Delta E $ is the energy uncertainty of the moving component. This
energy uncertainty $\Delta E $ is surely smaller than the total energy
$E=mc^2$ of the component itself. (Physically, this condition means that one
is sure that there is only  {\it one\/} clock; the case $\Delta E > E$
would mean that it is impossible to distinguish between a single clock, or
a clock plus a pair of clock-anticlock, or a component plus two
such pairs, etc.)  Furthermore, any clock provides information; therefore,
signals have to be able to leave it. To make this possible, the clock may not
be a black hole; its mass $m$ must therefore  be smaller than the
Schwarzschild mass for its size, {\it i.e.} $m \le c^2l/G$, where $l$ is the
size of the clock (here and in the following we neglect factors of order
unity). Finally, the size $l$ of the clock must be smaller than $c \,\delta t$
itself, to allow a sensible measurement of the time interval $\delta t$: one
has $l \le c \,\delta t$. (It is amusing to study how a clock larger than $c
\,\delta t$ stops being efficient, due to the loss of rigidity of its
components.) Putting all these conditions together one after the other, one
gets 
 $$ 
 \delta t \ge {\hbar G \over c^5 \delta t} \cv 
 $$ or
\begin{equation}
  \delta t \ge \sqrt{{\hbar G \over c^5 }} =  t_{\rm Pl}\cp 
\end{equation}
 In summary, from three simple properties of every clock, namely that one is 
sure to have only one of it, that one can read its dial, and that it gives
sensible readouts, one gets the general conclusion that {\it clocks cannot
measure time intervals shorter  than the Planck time}. 

Note that this argument is independent of the nature of the clock mechanism.
Whether the clock is powered by gravitational, electrical, plain mechanical or
even nuclear means, the relations still apply. Note also that gravitation is
essential in this argument.  A well-known study  on the limitations of clocks
due to their mass and their measuring time
has been published by Salecker and Wigner\cite{salek} and summarized in
pedagogical form by Zimmerman\cite{salek2}; the present  argument differs in
that it includes both quantum mechanics as well as gravity, and therefore
yields a different, lower, and much more fundamental limit.

The same result can also be found in other ways.\cite{gibbs} For example, any
clock small enough to measure small time intervals necessarily has a certain
energy spread, as described by quantum mechanics. Following general 
relativity, any energy density induces a deformation of space-time, and
signals from that region arrive with a certain delay. The energy uncertainty
of the source leads to a uncertainty in the delay. The expression from general
relativity\cite{wein}  for the deformation of the time part of the  line
element due to a mass $m$ is  $ \delta t = { m G / l c^3 }  $. If the mass
varies by $\Delta E/c^2$, the resulting uncertainty
$\Delta t$ in the delay, the external accuracy of the clock, is 
\begin{equation} 
 \Delta t = {\Delta E\; G \over l\, c^5} \cp
\end{equation} Now the energy uncertainty of the clock  is  bounded by the
uncertainty relation for time and energy, given the internal time  accuracy of
the clock. This internal accuracy must be smaller or equal than the  external
one. Putting this all together, one again finds the relation 
$\delta t \leq t_{\rm Pl}$. One is forced to conclude that {\it in nature
there is a minimum time  interval.} In other words,  {\it at Planck scales 
the term ``instant of time'' has no theoretical nor experimental backing.} It
therefore makes no sense to use it.

%-----------------------------------------------------------------------------
\subsection{Farewell to points in space}

In a similar way one can deduce that it is not possible to make a meter bar or
any other length measuring device that can measure lengths shorter than the
Planck length, as one can find already from $l_{\rm Pl}= c \,
t_{\rm Pl}$.\cite{mead} The straightforward way to measure the distance 
between two points is to put an object at rest at each position. In other
words, joint measurements of position and momentum are necessary for every
length measurement. Now the minimal length
$\delta l$ that can be measured is surely larger than the position
uncertainty  of the two objects. {}From the uncertainty principle it is known
that each object's position cannot be determined with a precision $\Delta l$
smaller than that given by
$\Delta l \,\Delta p =  \hbar$, where $\Delta p$ is the momentum uncertainty.
Requiring to have only one object at each end means  $\Delta p < mc$, which
gives 
\begin{equation}
 \delta l \ge \Delta l \ge {\hbar \over m c } \cp
\end{equation}
 Furthermore, the measurement  cannot be performed if signals
cannot leave the object: they may not
be black holes. Their masses must therefore be so small that their
Schwarzschild radius
$r_{{\rm S}}=  2 G m / c^2$ is smaller than the distance $\delta l$ separating
them. Dropping again the factor  of 2,  one gets
\begin{equation}
 \delta l \ge \sqrt{\hbar G \over c^3 } =  l_{{\rm Pl}} \cp 
\end{equation} 
 Another way to deduce this limit reverses the role of general relativity and
quantum theory. To measure the distance between two objects, one has to
localize the first object with respect to the other within a certain interval
$\Delta x$. This object thus possesses a momentum uncertainty $\Delta p
\geq\hbar/\Delta x$ and therefore possesses an energy uncertainty $\Delta E = 
c (c^2m^2+(\Delta p)^2)^{1/2}
\geq c \hbar/ \Delta x$. But general relativity shows that a small volume
filled with energy changes the curvature, and thus the metric of the
surrounding space.\cite{msw, wein} For the resulting distance change $\Delta
l$, compared to empty space, one  finds\cite{mead, town, jack,ahlu, 
gara,amel} the
expression
  $\Delta l \approx G \Delta E/ c^4 $. In short, if one localizes 
a first particle
in space with a precision $\Delta x$, the distance to a second particle is
known only with precision $\Delta l$. The minimum length $\delta l$ that can
be measured is obviously larger than each of the quantities; inserting the
expression for $\Delta E$, one finds again that the minimum measurable length
$\delta l$ is  given by the Planck length.

As a remark, the Planck length being the shortest possible length, it follows
that there can be no observations of quantum mechanical effects for situations
in which the corresponding de~Broglie or Compton wavelength would be even
smaller. This is one of the reasons why in everyday,
macroscopic situations, {\it e.g.} in car-car collisions,  
one never observes quantum interference
effects,
in opposition
to the case of  proton-proton collisions.
%
%, or
%why  one never finds embryo-antiembryo pair production.

In summary, from two simple properties common to all length measuring
devices, namely that they can be counted and that they can be read out, one
arrives at the conclusion that {\it lengths smaller than the Planck length
cannot be found in measurements}. Whatever the method used, whether lengths
are measured with a meter bar or by measuring time of flight of particles
between the end points: one cannot overcome this fundamental limit. It follows
that in its usual sense as entity without size, {\it the concept  of ``point
in space'' has no experimental backing}.  In the same way, the term ``event'',
being a combination of ``point in space'' and ``instant of time'', also loses
its meaningfulness for the description of nature.

These results are often summarized in the so-called generalized uncertainty
principle\cite{mead}
   \begin{equation} \Delta p \Delta x \geq \hbar/2 + f {G \over c^3} (\Delta
p)^2  \ \ , \label{uncertainty} 
 \end{equation}
 where $f$ is a numerical factor around unity. A similar expression holds for
the time-energy uncertainty relation. The first term on the right hand side is
the usual quantum mechanical one. The second term, negligible at everyday life
energies, plays a role only near Planck energies. It is due to the changes in
space-time induced by gravity at these high energies. One notes that the
generalized principle automatically implies that $\Delta x$ can never be
smaller than $f^{1/2} l_{\rm Pl}$. 

The generalized uncertainty
principle is derived in exactly the same way in which Heisenberg derived 
the original uncertainty principle $\Delta p \Delta x \geq \hbar/2$ for an
object: by using the deflection of light by the object under a microscope. 
A careful re-derivation of the process, not disregarding gravity, yields 
equation (\ref{uncertainty}).\cite{mead}

For this reason all approaches which try to unify quantum mechanics and gravity
must yield this relation; indeed it appears in canonical quantum 
gravity\cite{asht}, in superstring theory\cite{sst}, and in the quantum group
approach\cite{magg}, sometimes with an additional term proportional to
$(\Delta x)^2$ on the right hand side of equation
(\ref{uncertainty}).\cite{kempf}

Quantum mechanics starts when one realizes that the classical concept of
action makes no sense  below the value of
$\hbar$; similarly, unified theories such as quantum gravity start when one
realizes that the classical concepts of time and length make no sense near 
Planck values.  However, the usual
 description of space-time 
 does contain such small values; it claims  the existence of intervals
smaller than the smallest measurable one.  {\it Therefore the continuum
description of space-time has to be abolished in favor of a more appropriate
one.}

This is clearly expressed in a new uncertainty relation appearing at Planck
scales.\cite{hell} Inserting $ c\Delta p \geq \Delta E \geq \hbar/
\Delta t$ into equation (\ref{uncertainty}) one gets
\begin{equation}
 \Delta x \Delta t \geq { \hbar G / c^4} =  t_{\rm Pl} l_{\rm Pl}\  ,  
\end{equation}
 which of course has no counterpart in standard quantum mechanics. A final way
to  convince one-self that points have no meaning is that a point is an entity
with vanishing volume; however, the minimum volume possible in nature is the
Planck volume
$V_{\rm Pl}= l_{\rm Pl}^3$. 

Space-time points are idealizations of events. But this idealization is
incorrect. Using the concept of ``point'' is equivalent to the use of the
concept of ``ether'' a century ago: one cannot measure it, and it is useful to
describe observations only until one has found the way to describe them without
it.

%-----------------------------------------------------------------------------
\subsection{Farewell to the space-time manifold}

But the consequences of the Planck limits for time and space measurements can
be taken much further.   To put the previous results in a different way,
points in space and time have size, namely the Planck size. It is commonplace
to say that given any two points in space or two instants of time, there is
always a third in between. Physicists sloppily call this property continuity, 
mathematicians call it denseness. However, at Planck dimensions, this property
cannot hold since intervals smaller than the Planck time can never be found:
thus points and instants are not dense, and {\it between two points there is
not always a third.} But this means that {\it space and time are not
continuous.} Of course, at large scales they are  -- approximately --
continuous, in the same way that a piece of rubber or a liquid seems
continuous at everyday dimensions, but is not at small scales. This means that
to avoid Zeno's paradoxes resulting from the infinite divisibility of space
and time one does not need any more the system of differential calculus; one
can now directly dismiss the paradoxes because of wrong premises on the nature
of space and time.

But let us go on.  Special relativity, quantum mechanics and general
relativity all rely on the idea that time can be defined for all points of a
given reference frame. However, two clocks at a distance $l$ cannot be
synchronized with arbitrary precision.  Since the distance between two clocks
cannot be measured with an error smaller than the Planck length $l_{\rm Pl}$,
and transmission of signals is necessary for synchronization, it is not
possible to synchronize two clocks with a better precision than the time
$l_{{\rm Pl}}/c= t_{\rm Pl}$, the Planck time. Due to this impossibility to
synchronize clocks precisely, the idea of a single time coordinate for a whole
reference frame is only approximate, and cannot be maintained in a precise
description of nature.

Moreover, since the difference between events cannot be measured with a
precision better than a Planck time, for two events distant in time by this
order of magnitude,  it is not possible to say which one precedes the other
with hundred percent certainty. This is an important result. If events
cannot be ordered at Planck scales, the concept of time, which is introduced in
physics to describe sequences, cannot be defined at all. In other words, after
dropping the idea of a common time coordinate for a complete frame of
reference,  one is forced to drop the idea of time at a single ``point'' as
well. For example, the concept of `proper time' loses its sense at Planck
scales.

In the case of space, it is straightforward to use the same arguments to show
that length measurements do not allow us to speak of continuous space, but only
about approximately continuous space. Due to the lack of measurement precision
at Planck scales, the concept of spatial order, of translation
invariance and isotropy of the vacuum, and of global coordinate systems lack
experimental backing at those dimensions.

But there is more to come. The very existence of a minimum length contradicts
special relativity, where it is shown that whenever one changes to a moving
coordinate system a given length undergoes a Lorentz contraction. A minimum
length cannot exist in special relativity; therefore, {\it at Planck
dimensions, space-time is neither  Lorentz invariant, nor diffeomorphism
invariant, nor dilatation  invariant.} All the symmetries at the basis of
special and general relativity are thus only approximately valid at Planck
scales.

Due to the imprecision of measurement, most familiar concepts used to describe
spatial relations become useless. For example, the concept of {\it metric}
also loses its usefulness at Planck scales. Since distances cannot be 
measured with precision, the metric cannot be determined. One deduces that it
is impossible to say precisely whether space is flat or curved. In other
words, {\it the impossibility to measure lengths exactly is equivalent to
fluctuations of the curvature}.\cite{mead,cuflu}

In addition,  even the number of space dimensions makes no  sense at Planck
scales. Let us remind ourselves how one determines this number experimentally.
One possible way is to determine how many points one can choose in space such
that all their distances are equal. If one can find at most
$n$ such points, the space has $n-1$ dimensions.  One recognizes directly that
without reliable length measurements there is no way to determine reliably the
number of dimensions of a space at Planck scales with this method.

Another simple way to check for three dimensions is to make a knot  in a shoe
string and glue the ends together: if it stays knotted under all possible
deformations, the space has three dimensions. If it  can be unknotted, space
has more than three dimensions, since in such spaces knots do not exist.
Obviously, at Planck dimensions the measurement errors do not allow to say
whether a string is knotted or not, because at crossings one cannot say
which strand lies above the other. 

There are many other methods to determine the dimensionality of 
space.\cite{sdimen} All these methods use the fact that the concept of
dimensionality is based on a precise definition of the concept of
neighborhood. But at Planck scales, as just mentioned, length measurements do
not allow us to say with certainty whether  a given point is inside or outside
a given volume. In short, whatever method one uses, the lack of reliable length
measurements means that {\it  at Planck scales, the dimensionality of physical
space is not defined.} It should therefore not come as a surprise that when
{\it approaching\/} those scales,  
 one could get a scale-dependent answer, different from three.

The reason for the troubles with space-time become most evident when one
remembers the well-known definition by Euclid:\cite{euclid} ``A point is that
which has no part.'' As Euclid clearly understood, a physical point, and
here the stress is on {\it physical}, cannot be defined {\it without} some
measurement method. A physical point is an idealization of position, and as
such includes measurement right from the start. In mathematics however, Euclid's
definition is rejected, because mathematical points do not need metrics for
their definition. Mathematical points are elements of a set, usually called
a space. In mathematics, a measurable or a metric space is a set of points
equipped {\it in addition\/} with a measure or a metric. Mathematical points
do not need a metric for their definition; they are basic entities. In
contrast to the mathematical situation, the case of physical space-time, the
concepts of measure and of metric are {\it more fundamental} than that of a
point.  The difficulty of distinguishing physical and mathematical space
arises from the failure to distinguish a mathematical metric from a physical
length measurement. 

Perhaps the most beautiful way to make this point clear is the Banach-Tarski
theorem or paradox, which shows the limits of the concept of volume.\cite{BT}
This theorem states that a sphere made of {\it mathematical points} can be cut
into six pieces in such a way that two sets of three pieces can be put
together and form two spheres, each of the {\it same volume} as the original
one. However, the necessary cuts are ``infinitely'' curved and thin.  For
physical matter such as gold, unfortunately -- or fortunately -- the existence
of a minimum length, namely the atomic distance, makes it impossible to
perform such a cut. For vacuum, the puzzle  reappears: for example, the energy
of its zero-point fluctuations  is given by a density times the volume;
following the Banach-Tarski theorem, since the concept of volume is ill
defined, the total energy is so as well. However, this problem is solved by
the Planck length, which provides a fundamental length scale also for the
vacuum. 

In summary, {\it physical space-time cannot be a set of mathematical points.}
But the surprises are not finished.  At Planck dimensions, since both temporal
order  and spatial order break down, there is no way to say if the distance
between two near enough space-time regions is space-like or time-like. One
cannot distinguish the two cases. {\it At Planck scales, time and space cannot
be distinguished from each other}. In summary, space-time at Planck scales is
not continuous, not ordered, not endowed with a metric, not four-dimensional,
not made of points. If we  compare this with the definition of the term
manifold, (a manifold is what locally looks like an euclidean space)  not one
of its defining properties is fulfilled. We arrive at the conclusion that {\it
the concept of a space-time manifold has no backing at Planck scales}.
Even though both general relativity and quantum mechanics use continuous
space-time, the combination of both theories does not. This is one reason why
the idea is so slow to disappear.

%---------------------------------------------------------------------------
\subsection{Farewell to observables and measurements}

To complete this state of affairs,  if space and time are not continuous, all
quantities defined as derivatives versus space or time are not defined
precisely. Velocity, acceleration, momentum, energy, etc., are only defined in
the classical approximation of continuous space and time. Concepts such as
`derivative', `divergence-free', `source free', etc., lose their meaning at
Planck scales. Even the important tool of the evolution equation, based on
derivatives, such as the Schr\"odinger or the Dirac equation, cannot be used
any more.

In fact, all physical observables are defined using at least length and time
measurements, as is evident from any list of physical units. Any such table
shows that all physical units are products of powers of length, time (and
mass) units. (Even though in the SI system electrical quantities have a
separate base quantity, the ampere, the argument still holds; the ampere is
itself defined by measuring a force, which is measured using the three base
units length, time, and mass.) Now, since time and length are not continuous,
observables themselves are not continuously varying.  {\it This means that at
Planck scales, {observables} (or their components in a basis) are not to be
described by real numbers with -- potentially -- infinite precision.} 
Similarly, if time and space are not continuous, the usual expression for an
observable  quantity
$A$, namely $A(x,t)$, does not make sense: one has to find a more appropriate
description. {\it Physical fields cannot be described by continuous functions
at Planck scales.}

In quantum mechanics this means that it makes no sense to define
multiplication of observables by real numbers. Among others,
observables do not form a linear algebra. (One recognizes directly that due to
measurement errors, one cannot prove that observables do form such an
algebra.) This means that {\it observables are not described by operators at
Planck scales.} Moreover, the most important observables are the gauge
potentials. Since they do not  form an algebra, {\it gauge symmetry is not
valid at Planck scales}. Even innocuous looking expressions such as
$[x_i,x_j] = 0$ for $x_i\neq x_j$, which are at the basis of quantum field
theory, become meaningless at Planck scales. Even worse, also the
superposition principle cannot be backed up by experiment at those scales.

Similarly, permutation symmetry is based on the premise that one can
distinguish two points by their coordinates, and then exchange particles at 
those two locations. As just seen, this is not possible if the distance
between the two particles is small; one concludes that {\it permutation
symmetry has no experimental backing at Planck scales}.

Even discrete symmetries, like charge conjugation, space inversion,  and time
reversal   cannot be correct in that domain,  because there is no way  to
verify them exactly by measurement.  {\it CPT symmetry is not valid at Planck
scales.}

All these results are consistent: if there are no symmetries at Planck scales,
there also are no observables, since physical observables are representations
of symmetry groups. In fact, the limits on time and length measurements imply
that {\it the concept of measurement has no significance at Planck scales}.

%-----------------------------------------------------------------------------
\subsection{Can space-time be a lattice? Can it be dual?}

Discretizations of space-time have been studied already fifty years 
ago.\cite{schild} The idea that space-time is described as a lattice has also 
been studied in detail, for example by  Finkelstein\cite{fink} and by 't
Hooft.\cite{hooftlat} It is generally agreed that in order to get an
isotropic  and homogeneous situation for large, everyday scales, the lattice
cannot be periodic, but must be random.\cite{bombposet}  Moreover any fixed
lattice violates the result that there are no lengths smaller than the Planck
length: due to the Lorentz contraction, any moving observer  would find
lattice distances smaller than the Planck value. 

If space-time is not a set of points or events, it must be a set of something
else. Three hints already appear at this stage. The first necessary step to
improve the description of motion starts with the recognition that to abandon
``points'' means to abandon the \ii{local} description of physics. Both
quantum mechanics and general relativity  assume that the  phrase `observable
at a point' had a precise meaning. Due to the impossibility of describing
space as a manifold, this expression is no longer useful.  The
unification of general relativity and quantum physics forces a  {\it
non-local\/} description of nature at Planck scales.

The existence of a minimal
length implies that there is no way to physically distinguish locations that
are spaced by even smaller distances. One is tempted to conclude that
therefore {\it any} pair of locations cannot be distinguished, even if they 
are one
meter apart, since on any path joining two points, any two nearby locations
cannot be distinguished. One notices that this situation is similar to the
question on the size of  a cloud or that of an atom. Measuring water density
or electron  density, one finds  non-vanishing values at any distance from the
center; however, an effective size can still be defined, because it is very
improbable to see effects of a cloud's or of an atom's  presence at distances
much larger than this effective size. Similarly, one guesses that space-time
points at macroscopic distances can be distinguished because the probability
that they will be confused drops rapidly with increasing distance. In short,
one is thus led to a {\it probabilistic\/} description of space-time. It
becomes a macroscopic observable, the \ii{statistical, or thermodynamic limit} 
of some microscopic entities.

One notes that a {\it fluctuating} structure for space-time would also 
avoid the problems of fixed structures with Lorentz invariance.  This 
property is of course compatible with a statistical description.
In summary, the experimental observations of Lorentz invariance, isotropy, and
homogeneity, together with that of a minimum distance, point towards a
fluctuating  description of space-time. In the meantime, research
efforts in quantum gravity, superstring theory and in quantum groups have
confirmed independently from each other that a probabilistic description of
space-time, together with a {\it non-local} description of observables at
Planck dimensions, indeed resolves the contradictions between general
relativity and quantum theory. To clarify the issue, one has to turn to the
concept of particle.

Before that, a few remarks on one of the most important topics in 
theoretical physics at present: duality.  String theory is build around a
new supposed symmetry {\it space-time duality} (and its  generalizations),
which states that in nature, physical systems of  size $R$ are
indistinguishable from those of size $l_{{\rm  Pl}}^{2}/R$.  (In natural
units this symmetry is often written $R 
\leftrightarrow 1/R$.)  Since this relation implies that there is a  symmetry
between systems larger and smaller that the Planck length,  it is in contrast
with the result above that lengths smaller than the  Planck length do not
make sense. Physical space-time therefore is not dual; but since present
string theory includes the assumption that intervals of any size can be used
to describe nature, it needs the duality symmetry to filter out the thus
introduced unphysical situations. Duality does this by explaining that any
system which would have smaller dimensions than the Planck length is in fact
a system larger than this length.

%---------------------------------------------------------------------------
\subsection{Farewell to particles}

Apart from space and time, in every example of motion, there is some object
involved. One of the important discoveries of the natural sciences was that
all objects are made of small constituents, called
\ii{elementary particles}.
Quantum theory shows that all composite, non-elementary objects have a simple
property: they have a finite, non-vanishing size. This property allows us to
determine whether a particle is elementary or not. If  it behaves like a point
particle, it is elementary. At present, only the leptons (electron, muon, tau
and the neutrinos), the quarks, and the radiation quanta of the
electromagnetic, the weak and the strong nuclear interaction (the photon, the
W and Z bosons, the gluons) have been found to be
elementary.\cite{elementary}  A few more elementary particles are predicted by
various refinements of the standard model. Protons, atoms, molecules, cheese,
people, galaxies, etc., are all composite (see table
\ref{tsch}). Elementary particles are characterized by their vanishing size,
their spin, and their mass.

The {\it size} of an object, e.g. the one given in table 1, is defined as the
length at which one observes differences from point-like behavior. This is
the way in which, using alpha particle scattering, the radius of the atomic
nucleus was determined for the first time  in Rutherford's experiment.

Speaking simply, the size $d$ of an object is determined by measuring the
interference pattern in the scattering of a beam of probe particles. This is
the way one determines sizes of objects when one looks at them, using scattered
photons. In order to observe such interference effects,  the
wavelength
$\lambda$ of the probe must be smaller than the object size $d$ to be
determined. The de~Broglie wavelength of the probe is given by the mass and
the relative velocity $v$ between the probe and the unknown object, {\it i.e.}
one needs 
$d >\lambda =  \hbar / (mv ) \geq \hbar/ (mc) $. On the other hand, in order
to make a scattering experiment possible, the object must not be a black hole,
since then it would simply swallow the infalling particle.  This means that
its mass $m$ must be smaller than that of a black hole of its size, i.e., from
equation (\ref{SCHW}), that
$ m < d c^2/G $; together with the previous condition one gets 
\begin{equation} 
 d > \sqrt{{ \hbar G \over c^3 }} =  l_{\rm Pl} \cp 
\end{equation}
   In other words, there is no way to observe that an object is smaller than
the Planck length. {\it There is thus no way in principle to deduce from
observations that a particle is point-like.} In fact, it makes no sense to use
the term ``point particle'' at all. Of course the existence of a minimal
length both for empty space and for objects, are related. If the
term ``point of space'' is meaningless, then the term ``point particle'' is so
as well. Note that as in the case of time, the lower limit on length results
from the combination of quantum mechanics and general relativity. 
(Note also that the minimal size of a particle has nothing to do with the
impossibility, quantum theory, to localize a particle to within better than
its Compton wavelength.)

The size $d$ of  any elementary particle, which following the
conventional quantum field description is  zero, is surely smaller than its
own Compton wavelength $\hbar/(mc)$. Moreover, we have seen above that a
particle's size is always larger than the Planck length: $d > l_{\rm Pl}$.
Eliminating the size $d$ one gets a condition for the mass $m$  of any
elementary particle, namely 
\begin{equation} m < {\hbar \over c \, l_{\rm Pl} } =  \sqrt{{\hbar c \over
G}} =  m_{\rm Pl}= \csd{2.2
\cstimes 10^{-8}}{kg} = \csd{1.2 \cstimes  10^{19}}{GeV/c^{2}}
\end{equation}  (This limit, the so-called \ii{Planck mass}, corresponds
roughly to the mass of a ten days old human embryo, or, equivalently that of a
small flea.) In short, {\it the  mass of any elementary particle must be
smaller than the Planck mass.} This fact is already mentioned as
``well-known'' by Sakharov\cite{sakha} who explains  that these hypothetical
particles are sometimes called `maximons'. And indeed, the known elementary
particles all have masses well below the Planck mass. (Actually, the question
why their masses are so incredibly much smaller than  the Planck mass is one
of the main questions of high energy physics. But this is another story.)
%\cite{generation} 

There are many other ways to arrive at this mass limit. For example, in order
to measure mass by scattering -- and that is the only way for very small
objects -- the Compton wavelength of the scatterer must be larger than the
Schwarzschild radius; otherwise  the probe would be swallowed.  Inserting the
definition of the two quantities and neglecting the factor 2, one gets again
the limit $ m <  m_{{\rm Pl}} $. (In fact it is a general property of 
descriptions of nature that  a minimum space-time interval leads to an upper
limit for elementary particle masses.\cite{wolf}) The importance of the
Planck mass will become clear shortly.

Another property connected with the size of a particle is its 
electric dipole moment. The stand model of elementary particles gives 
as {\it upper} limit for the dipole moment of the electron $d_{e}$ 
a value of\cite{bernreu}
\begin{equation} |d_{e}|  <  
\csd{10^{-39}}{m}\ e 
\end{equation} 
where $e$ is the charge of the electron. This value is 
ten thousand times smaller than $l_{{\rm Pl}}\ e$; using the fact 
that the Planck 
length is the smallest possible length, this implies either that charge
can be distributed in space, or that estimate is wrong, or 
that the standard model is wrong, or several of these. Only future will tell.

Let us return to some other strange consequences for particles. In quantum
field theory, the difference  between a virtual and a real particle is that a
real particle is on shell, {\it i.e.} it
obeys $E^2= m^2c^4 + p^2c^2$, whereas a virtual particle is off shell, i.e.
$E^2\neq m^2c^4 + p^2c^2$. Due to the fundamental limits of measurement
precision,  {\it  at Planck scales one cannot determine whether a particle is
real or virtual.}

But that is not all. Since antimatter can be described as matter moving
backwards in time, and since the difference between  backwards and forwards
cannot be determined  at Planck scales, \ii{matter and antimatter cannot be
distinguished at Planck scales.}

Particles are also characterized by their spin.  Spin describes two  properties
of a particle: its behavior under rotations (and if the particle is charged,
the behavior in magnetic fields) and its behavior under particle exchange.
The wave function of spin 1 particles  remain invariant under rotation of $2
\pi$, whereas  that of spin 1/2 particles changes sign. Similarly, the
combined wave  function of two spin 1 particles does not change sign under
exchange of particles,  whereas for two spin 1/2 particles it does.

One sees directly that both transformations are impossible to study at 
Planck scales.  Given the limit on position measurements, the position 
of the axis of a rotation cannot be well defined, and rotations become 
impossible to distinguish from translations.  Similarly, position 
imprecision makes the determination of precise separate positions for 
exchange experiments impossible.  In short, {\it spin cannot be 
defined at Planck scales, and fermions cannot distinguished from 
bosons, or, differently phrased, matter cannot be distinguished from 
radiation at Planack scales.} One can thus easily imagine that 
supersymmetry, a unifying symmetry between bosons and fermions, 
somehow must become important at those dimensions.  Let us now move to 
the main property of elementary particles.

%---------------------------------------------------------------------------
\subsection{Farewell to mass}

The Planck mass divided by the Planck volume, {\it i.e.} the Planck density,
is  given by \begin{equation}
 \rho_{\rm Pl}= {c^5 \over G^2 \hbar} = \csd{5.2 \cstimes 10^{96}}{kg/m^3}
\end{equation}
  and is a useful concept in the following. If one wants to measure the
(gravitational) mass $M$ enclosed in a sphere of size  $R$ and thus (roughly)
of volume
$R^3$, one way to do this is to put a test particle  in orbit around
it at a distance $R$. The universal law of gravity then gives for the mass $M$
the expression $M=R v^2 / G$, where $v$ is the speed of the orbiting test
particle. {}From
$v < c$, one thus deduces that
$M< c^2 R/ G$;  using the fact that the minimum value for $R$ is the Planck
distance,  one gets (neglecting again factors of order unity)  a limit for the
mass density, namely  
\begin{equation}
   \rho < \rho_{\rm Pl} \cp
\end{equation} 
 In other words, {\it the Planck density is the maximum possible value for
mass density.} In particular, in a volume of Planck dimensions,  one
cannot have a larger mass than the Planck mass.

Interesting things happen when one starts to determine the error $\Delta M$ of
the mass measurement in a Planck volume. Let us return to the mass 
measurement by an orbiting probe.  {}From the relation $ GM = r v^2$ one
follows $ G \Delta M = v^2 \Delta r + 2 v r \Delta v > 2 v r
\Delta v = 2 GM \Delta v /v$. For the error $\Delta v$ in the velocity
measurement one has the uncertainty relation $\Delta v  \geq \hbar /(m
\Delta r) + \hbar / (M R) \geq \hbar / (M R)$.  Inserting this in the previous
inequality, and forgetting again the factor of 2, one gets that the mass
measurement error $\Delta M$ of a mass $M$  enclosed in a volume of size
$R$  follows 
 \begin{equation} 
   \Delta M \geq {\hbar \over c R} \ \cp
  \label{bal} 
 \end{equation}
 Note that for everyday situations, this error is extremely small,
 and other errors, such as the technical limits of the balance, are much
larger.

As a check, let us take another situation, and use relativistic expressions,
in order to show that the result  still holds. Imagine having  a mass $M$ in a
box of size $R$ and weighing the box.  (It is supposed  that either the box is
massless, or that its mass is subtracted by the scale.) The mass error is
given by $\Delta M = \Delta E / c^2 $, where $\Delta E$ is due to the
uncertainty in kinetic energy of the mass inside the box.
 Using the expression $E^2 = m^2c^4 + p^2 c^2$ one gets that $\Delta M \geq
\Delta p / c$, which again reduces to equation (\ref{bal}). 

For a box of Planck dimensions, the mass measurement {\it error} is given
by the Planck mass. But from above  we know that the mass  that can be put
inside such a box is itself not larger than the Planck mass.  In other words,
the mass measurement {\it error} is larger (or at best equal) to the mass
contained in a  box of Planck dimension. In other words,  if one builds a
balance with two boxes  of Planck size, one empty and the other full, as shown
in the figure, nature cannot decide which way the balance should hang! Note
that even a repeated or a continuous measurement would not resolve the
situation: the balance would then randomly change inclination.

The same surprising answer is found if instead of measuring the gravitational
mass one measures the inertial mass. The inertial mass for a small object is
determined by touching it, {\it i.e.}, physically speaking, by performing a
scattering  experiment. To determine the inertial mass inside a region of
size  $R$, a probe  must have a wavelength smaller that $R$, and thus a
corresponding high energy.  A high energy means that the probe also attracts
the particle through gravity. (One thus finds the intermediate result that
{\it at Planck scales, inertial and gravitational mass cannot be
distinguished.} Also the balance experiment shown in the figure  makes this
point: at Planck scales, the two effects of mass are always inextricably
linked.) In any scattering experiment, {\it e.g.}  in a Compton-type
experiment, the mass measurement is performed by  measuring the wavelength of
the probe before and after the  scattering experiment. We know from above that
there always  is a minimal wavelength error given by the Planck length
$l_{\rm PL}$. In order to determine the mass in a  Planck volume, the probe
has to have a wavelength of the Planck length.  In other words, the mass error
is as large as the Planck mass itself: $\Delta M \geq M_{\rm Pl} $. This limit
is thus a direct consequence of the limit on length and space  measurements.

This result has an astonishing consequence. In these examples, the measurement
error is independent of the mass of the scatterer, i.e. independent of whether
one starts with a situation in which there is a particle in the original
volume, or if there is none.  One thus finds that in a volume of Planck size,
it is impossible to say if there is one particle or none when weighing it or
probing it with a beam!  In short,
\ii{vacuum, {\it i.e.} empty space-time can not be distinguished from matter
at Planck scales.} Another, often used  way to express this is to say that
when a particle of Planck energy travels  through space it will be scattered
by the fluctuations of space-time itself, making it thus impossible to say
whether it was scattered by empty space-time or by matter.   These surprising
results stem from the fact that whatever definition of mass one uses, it is
always measured via length and  time measurements. (This is even so for
normal  weight scales: mass is measured by the displacement of some part of
the machine.) And the error in these measurements makes it impossible to
distinguish vacuum from matter.

To put it another way, if one measures the mass of a piece of vacuum of size
$R$, the result is always at least $\hbar/ c R$; there is no possible way to
find a perfect vacuum in an experiment. On the other hand, if one measures the
mass of a particle, one finds that the result is size dependent; at Planck
dimensions it approaches the Planck mass for every type of elementary particle.
This result is valid both for massive and massless particles.

Using another image, when two particles are approached to lengths of the order
of the Planck length, the error in the length measurements makes it impossible
to say whether there is something or nothing between the two objects.  In
short, \ii{matter and vacuum get mixed-up at Planck dimensions.} This is an
important result: since  both mass and empty space-time can be mixed-up, one
has confirmed that they are made of the same ``fabric'', confirming the idea
presented at the beginning.  This idea is now commonplace in all attempts to
find a unified description of nature.

This approach is corroborated by the attempts of quantum mechanics
in highly curved space-time,  where a clear distinction between the vacuum and
particles is not possible; the well-known Unruh radiation\cite{unruh}, namely
that any observer either accelerated in vacuum or in a gravitational field in
vacuum still detects particles  hitting him, is one of the examples which
shows that for curved space-time  the idea of vacuum as a particle-free space
does not work. Since at Planck scales it is impossible to say whether space is
flat or not, it follows that it is also impossible to say whether it contains 
particles or not.

In summary: the usual concepts of matter and of radiatiation are not 
applicable at Planck dimensions.  Usually, it is assumed that matter 
and radiation are made of interacting elementary particles.  The 
concept of an elementary particle is that of an entity which is 
countable, point-like, real and not virtual, with a definite mass, a 
definite spin, distinct from its antiparticle, and most of all, 
distinct from vacuum, which is assumed to have zero mass.  All these 
properties are found to be incorrect at Planck scales.  One is forced 
to conclude that {\it at Planck dimensions, it does not make sense to 
use the concepts of `mass', `vacuum', `elementary particle', 
`radiation', and `matter'.}

One then can ask at which dimension particles and vacuum {\it can} be
distinguished. This is obviously possible at everyday dimensions, e.g.~of the
order of \csd{1}{m}, and is experimentally possible up to \csd {10^{-18}}{m}.
Present ideas in particle physics suggest that particles are defined at least
until the scale of unification of the electromagnetic, weak and strong
interaction, i.e.~until about \csd{10^{-31}}{m}. Somewhere
between there and the Planck
length of \csd{10^{-35}}{m} the distinction between particles and space-time
becomes impossible, perhaps in some gradual way. 

%-----------------------------------------------------------------------------
\subsection{Unification}

In this rapid walk, one has thus destroyed all the experimental pillars of
quantum theory: the superposition principle, space-time symmetry, gauge
symmetry, and the discrete symmetries. One also has destroyed the foundations
of general relativity, namely the existence of a space-time manifold, of the
field concept, the particle concept, and of the concept of mass. It was even
shown that matter and space-time  cannot be distinguished. It seems that one
has destroyed every concept used for the description of motion,  and thus
made the description of nature impossible. One naturally asks whether
one can save the situation.

To answer this question, one needs to see what one has gained from 
this sequence of destructive arguments.  First, since it was found 
that matter is not distinguishable for vacuum, and since this is 
correct for all types of particles, be they matter or radiation, one 
has a clear argument to show that the quest for unification in the 
description of elementary particles is correct and necessary.

Moreover, since the concepts `mass', `time', and `space' cannot be 
distinguished from each other, we also know that a new, {\it single} 
entity is necessary to define both particles and space-time.  To find 
out more about this new entity, three approaches are being pursued in 
present research.  The first, quantum gravity, especially the one 
using the Ashtekar's new variables and the loop 
representation\cite{ashtall}, starts by generalizing space-time 
symmetry.  The second, string theory\cite{sstall}, starts by 
generalizing gauge symmetries and interactions, and the third, the 
algebraic quantum group approach, looks for generalized permutation 
symmetries.\cite{maggall}

The basic result of all these descriptions is that what is usually called
matter and vacuum are two different aspects of one same ``soup'' of
constituents.  In this view, mass and spin are not intrinsic properties, but
are  properties emerging from certain configurations of the fundamental soup.
The fundamental constituents themselves have no mass, no spin, and no charge.
The three mentioned approaches to a unified theory have one thing in common:
the fundamental constituents are {\it extended}.
(In this way the non-locality required above is realized.) In a subsequent
article we will give a collection of simple arguments for this result.

The present speculations describe the vacuum as a fluctuating tangle of
extended entities and describe particles as knots or possibly even more complex
structures. Such models explain why space has three dimensions, give hope to
calculate the masses of particles, to determine their spin, and naturally
confirm that vacuum has no energy at large scales, but high energy at small
dimensions. At everyday  energies and length scales, the tangle is not
visible, the knots become point-like, and one recovers the smooth, empty
vacuum and the massive, spinning, charged and pointlike particles.  In this
way, these speculative models avoid the circular definition of particles and
space-time mentioned at the beginning. However, the details of this programme
are still hidden in the impenetrable, but fascinating future of physics.

%-----------------------------------------------------------------------------
\subsection{Acknowledgement}

This work resulted from an intense discussion exchange with Luca Bombelli, now
at the Department of Physics and Astronomy of the University of Mississippi.

%---------------------------------------------------------------------------
%---------------------------------------------------------------------------

%---------------------------------------------------------------------------
%---------------------------------------------------------------------------
\newpage

\subsection*{Table}

{\ }

\vskip 10mm

\begin{table}[ht]
\hbox{\hss
\begin{tabular}{\zz l\zz |\zzs l\zz |\zzs l\zz |\zzs l\zz |\zzs l\zz |\zzs
l\zz |\zzs l\zz }
       & size:        & mass  & Schwarz- & ratio      & Compton & ratio \\
 Object& diameter & $m$ & schild & $d/r_{\rm S}$ & wave &
 $d/\lambda_{\rm C}$\\
       &  $d\,$&           &radius $r_{\rm S}$ &   & length
 $\lambda_{\rm C}$ & \cstabhlinedown \\
 \hline \cstabhlineup
 galaxy & \csd{\approx 1}{Zm}& \csd{\approx 5 \cstimes 10^{40}}{kg}&
 \csd{\approx 70}{Tm} & \csd{\approx 10^7}{} & \csd{\approx10^{-83}}{m} &
 \csd{\approx10^{104}}{}
 \\
 neutron star &\csd{10}{km}  &  \csd{2.8 \cstimes 10^{30}}{kg}& \csd{4.2}{km}
 & \csd{2.4}{} & \csd{1.3 \cstimes 10^{-73}}{m}& \csd{8.0 \cstimes 10^{76}}{}
 \\      
 sun    &\csd{1.4}{Gm}  &  \csd{2.0 \cstimes 10^{30}}{kg}& \csd{3.0}{km}
 & \csd{4.8 \cstimes 10^5}{} & \csd{1.0 \cstimes 10^{-73}}{m} & 
 \csd{8.0 \cstimes 10^{81}}{}
 \\
 earth &\csd{13}{Mm} & \csd{6.0 \cstimes 10^{24}}{kg} & \csd{8.9}{mm}
 & \csd{1.4 \cstimes 10^9}{} & \csd{5.8 \cstimes 10^{-68}}{m} & 
 \csd{2.2 \cstimes 10^{74}}{}
 \\
 human & \csd{1.8}{m} & \csd{75}{kg} & \csd{0.11}{ym} & 
 \csd{1.6 \cstimes 10^{25}}{}& \csd{4.7 \cstimes 10^{-45}}{m} &
 \csd{3.8 \cstimes 10^{44}}{}
 \\
 molecule & \csd{10}{nm} & \csd{0.57}{zg} &  \csd{8.5 \cstimes 10^{-52}}{m} & 
 \csd{1.2 \cstimes 10^{43}}{}& \csd{6.2 \cstimes 10^{-19}}{m} &
 \csd{1.6 \cstimes 10^{10}}{}
 \\
 atom (${}^{12}$C)& \csd{0.6}{nm}  & \csd{20}{yg} &  \csd{3.0 \cstimes
10^{-53}}{m} & 
 \csd{2.0 \cstimes 10^{43}}{}& \csd{1.8 \cstimes 10^{-17}}{m} &
 \csd{3.2 \cstimes 10^{7}}{}
 \\
 proton p& \csd{2}{fm} & \csd{1.7}{yg} & \csd{2.5\cstimes10^{-54}}{m}&  
 \csd{8.0 \cstimes 10^{38}}{}& \csd{2.0 \cstimes 10^{-16}}{m} &
 \csd{9.6 }{}
 \\
 pion $\pi$& \csd{2}{fm} & \csd{0.24}{yg} & \csd{3.6\cstimes10^{-55}}{m}&  
 \csd{5.6 \cstimes 10^{39}}{}& \csd{1.5 \cstimes 10^{-15}}{m} &
 \csd{1.4 }{}
 \\
 up-quark u& \csd{<0.1}{fm} & \csd{0.6}{yg} &  \csd{9.0\cstimes10^{-55}}{m}&
 \csd{<1.1 \cstimes 10^{38}}{}& \csd{5.5 \cstimes 10^{-16}}{m} &
 \csd{< 0.18}{}
 \\
 electron e& \csd{<4}{am} & \csd{9.1\cstimes10^{-31}}{kg}  &
 \csd{1.4\cstimes10^{-57}}{m}  &  
 \csd{3.0 \cstimes 10^{39}}{}& \csd{3.8 \cstimes 10^{-13}}{m} &
 \csd{< 1.0 \cstimes 10^{-5}}{}
 \\
 neutrino $\nu_{\rm e}$&\csd{<4}{am} &\csd{<3.0\cstimes10^{-35}}{kg} &
 \csd{<4.5 \cstimes10^{-62}}{m}  &  
 \csd{n.a.}{}& \csd{> 1.1 \cstimes 10^{-8}}{m} &
 \csd{< 3.4 \cstimes 10^{-10}}{}
 \\
\end{tabular}
\hss} % end of hbox for centering

\bigskip
\caption{ The size, Schwarzschild radius, and Compton wavelength 
   of some objects appearing in nature. %\protect\cite{si} 
   A short reminder of the new SI prefixes: f:~$10^{-15}$, a:~$10^{-18}$,
   z:~$10^{-21}$, y:~$10^{-24}$,
   P:~$10^{15}$, E:~$10^{18}$, Z:~$10^{21}$, Y:~$10^{24}$. }

\protect\label{tsch}
\end{table}

%---------------------------------------------------------------------------
%---------------------------------------------------------------------------
%\newpage
\vskip 30mm

\subsection*{Figure caption}

{\ }

\vskip 10mm

Figure 1: A {\it Gedankenexperiment\/}
 showing that at Planck scales, matter and  vacuum cannot be distinguished.

\message{............................................................}
\message{A sketch of the figure can also be printed using Latex alone}
\message{by looking for this text and deleting the subsequent /comment .}
\message{............................................................}

\comment{  % take this away to have the raw figure printed
%---------------------------------------------------------------------------
%---------------------------------------------------------------------------
\newpage

\thispagestyle{empty}

{\ }

\vskip 1cm

\begin{verbatim}

A sketch of the figure, representing a balance with two 
boxes of size R, one of which contains a mass M.










                        o
             ___________|__________         
            /\                    /\        
           /  \                  /  \       
          /    \                /    \       
         /------\              /------\        
        / |  M | \            / |    | \      
       /  |  o |  \          /  |    |  \     
        \ ------ /            \ ------ /      
                                                       
          <-R-->                <-R-->       
   
\end{verbatim}
} %end of comment

% end of \comment

%---------------------------------------------------------------------------
%---------------------------------------------------------------------------
%---------------------------------------------------------------------------

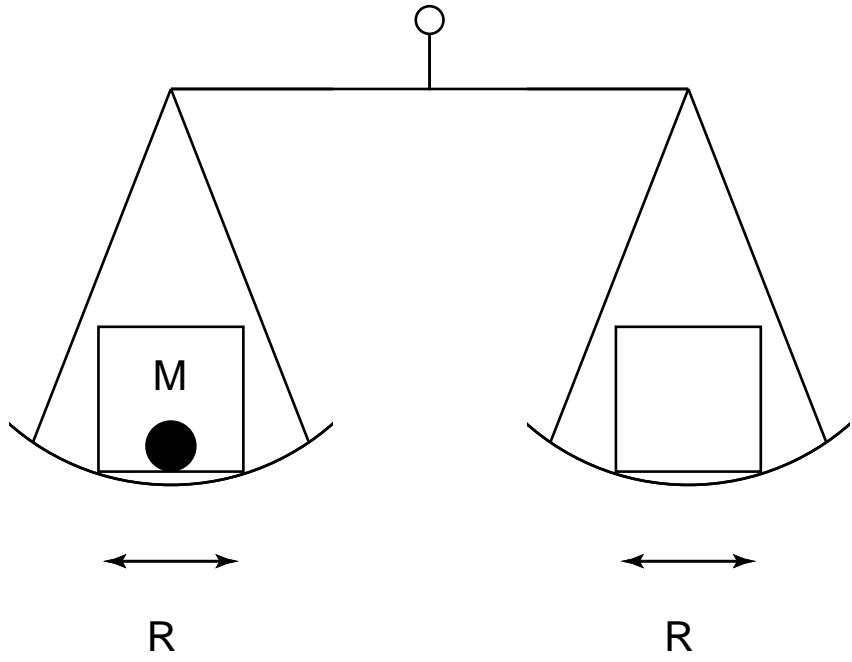
\end{document}